\documentclass[9pt]{article}
\usepackage{mathtools}
\usepackage{authblk}
\title{Low-error and broadband microwave frequency measurement in a silicon chip}

\author[1]{Mattia~Pagani}
\author[1]{Blair~Morrison}
\author[1]{Yanbing~Zhang}
\author[1]{Alvaro~Casas-Bedoya}
\author[2]{Timo~Aalto}
\author[2]{Mikko~Harjanne}
\author[2]{Markku~ Kapulainen}
\author[1]{Benjamin~J.~Eggleton}
\author[1]{David~Marpaung\thanks{Corresponding author: d.marpaung@physics.usyd.edu.au}}
\affil[1]{Centre for Ultrahigh Bandwidth Devices for Optical Systems (CUDOS), the Institute of Photonics and Optical Sciences (IPOS), School~of~Physics, University of Sydney, NSW 2006, Australia}
\affil[2]{VTT Technical Research Centre of Finland, Espoo, 02040, Finland}

\date{}

\begin{document}

\maketitle

\begin{abstract}
	Instantaneous frequency measurement (IFM) of microwave signals is a fundamental functionality for applications ranging from electronic warfare to biomedical technology. Photonic techniques, and nonlinear optical interactions in particular, have the potential to broaden the frequency measurement range beyond the limits of electronic IFM systems. The key lies in efficiently harnessing optical mixing in an integrated nonlinear platform, with low losses. In this work, we exploit the low loss of a 35 cm long, thick silicon waveguide, to efficiently harness Kerr nonlinearity, and demonstrate the first on-chip four-wave mixing (FWM) based IFM system. We achieve a large 40 GHz measurement bandwidth and record-low measurement error. Finally, we discuss the future prospect of integrating the whole IFM system on a silicon chip to enable the first reconfigurable, broadband IFM receiver with low-latency.
\end{abstract}

\section{Introduction}
The ability to measure the frequency of an unknown radio frequency (RF) or microwave signal, without relying on expensive spectrum analyzers and mixers, is a basic requirement for the development and testing of wireless systems. In particular, the ability to do so in real-time ($< 50$ ns) and with high accuracy ($< 1$\% error) is crucial for radar receivers used in electronic warfare (EW) \cite{ew_book} and biomedical technology \cite{patent}. This is known as instantaneous frequency measurement (IFM), and it involves mapping the signal frequency to a more easily measurable quantity, such as power.

The use of photonics for microwave IFM is attractive due to the possibility of pushing the frequency measurement range far beyond the capacity of electronic-based IFM systems \cite{capmany:07, capmany:14}. Since its conception in 2006 \cite{nguyen:06}, research into photonic-based IFM systems has progressed by various means \cite{emami:08, bui:09, drummond:09, yao:10, li:12}, and has recently culminated with the demonstration of a number of on-chip IFM systems \cite{marpaung:13_2, fandino:13, liu:15, burla:15}. This trend towards integration is especially important for IFM receivers, where the IFM delay is directly related to the length of the path traveled by the light. The inherently reduced size of integrated systems can thus lead to sub-nanosecond latency, as well as improved robustness and lower footprint \cite{marpaung:13}.

Nevertheless, most on-chip IFM demonstrations to date achieved measurement bandwidths below 10 GHz \cite{marpaung:13_2, fandino:13, liu:15}, lower than what is currently possible using electronic IFM systems. This limit was due to the trade-off between quality factor and free spectral range of the ring resonators employed by these prior demonstrations. A different technique made use of a highly compact on-chip Bragg grating filter with promising results, but a relatively high 2\% measurement error \cite{burla:15}. There are not many techniques that can simultaneously achieve wideband operation and low, sub-1\% measurement error. One such technique relies on mapping the RF signal frequency to the power of an optical idler generated through efficient nonlinear optical mixing, in a low-loss platform \cite{bui:09}. The stringent requirements on the nonlinear medium have meant that so far, all implementations \cite{bui:09, bui:13} were bound to using hundreds of meters of fiber, introducing high measurement delays. For high-bandwidth IFM, where low latency is key, it is crucial to harness the nonlinearity efficiently in a low-loss, short platform.

In this work, we report the first low-latency IFM system using on-chip four-wave mixing (FWM), and demonstrate 40 GHz measurement bandwidth, limited only by the measurement equipment, with record-low 0.8\% error. These ground-breaking results were enabled by the ultra-low loss and efficient FWM in a unique platform: a 35 cm long thick silicon waveguide with a highly compact footprint. This allowed us to generate an idler with enhanced signal-to-noise ratio, thereby significantly reducing the measurement error below the level demonstrated in fiber-based systems \cite{bui:09}. This is a breakthrough for high-performance on-chip IFM systems, and reveals a new and unique material platform for nonlinear optical processing of microwave signals \cite{marpaung:14}. Finally, we discuss the feasibility of integrating the whole IFM system onto a silicon chip, highlighting the potential for implementing the first, ultra-low latency, fully reconfigurable IFM system.

\section{Principle of operation}
\begin{figure}[htbp]
	\centering
	\includegraphics[scale=1]{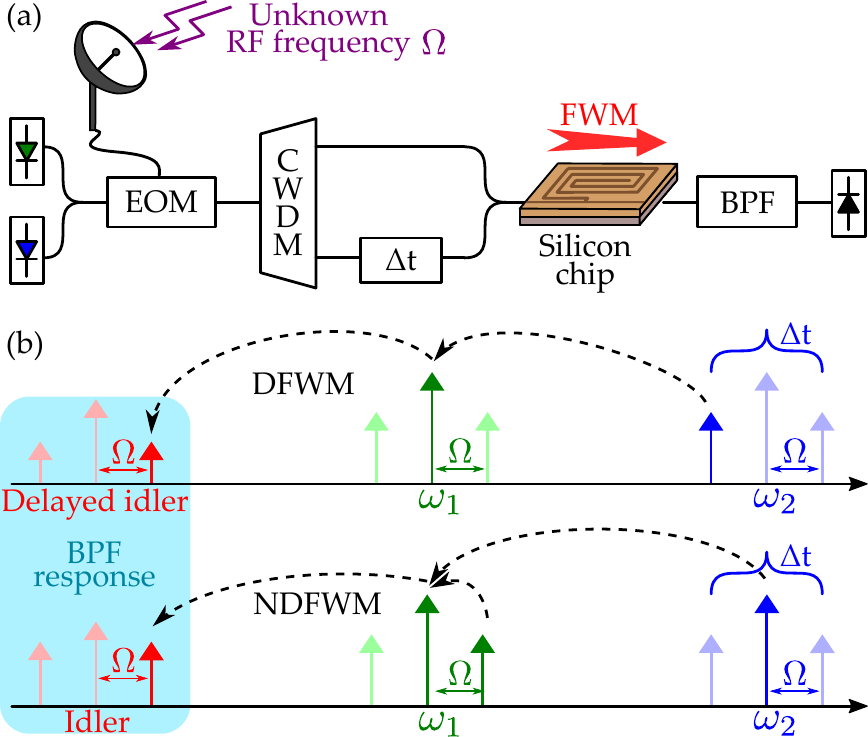}
	\caption{(a) Schematic of the on-chip FWM-based IFM system. EOM: electro-optic modulator; CWDM: coarse wavelength division multiplexer; $\Delta t$: tunable delay element; BPF: optical bandpass filter. (b) Degenerate (DFWM) and non-degenerate (NDFWM) mixing processes between the two channels.}
	\label{fig_principle}
\end{figure}

The basic structure of our FWM-based IFM system is shown in Fig. \ref{fig_principle}(a). An RF signal of unknown frequency $\Omega$ is received and modulates two CW optical carriers at different frequencies $\omega_1$, $\omega_2$. This results in two copies of an optical signal, in two different channels. These two optical signals are then demultiplexed using a coarse wavelength division multiplexer (CWDM). In this way, one signal is delayed by $\Delta t$ relative to the other, before both are recombined. The two signals are then sent through a nonlinear medium, where they mix through FWM, generating idlers in adjacent channels.

There are three idler components generated in the vicinity of frequency $2\omega_1 - \omega_2$. The middle component is an idler ``carrier'', resulting from degenerate FWM (DFWM) between the two original carriers. The two remaining components, or idler sidebands, occur at frequencies $2\omega_1-\omega_2\pm\Omega$. There are in fact two FWM processes which occur simultaneously, shown in Fig. \ref{fig_principle}(b), giving rise to two separate pairs of idler sidebands. The first is a DFWM process, which gives rise to a delayed pair of idler sidebands. Neglecting dispersion, the complex field for the upper DFWM idler sideband is
\begin{equation}
	E_\text{DFWM}(t) \propto 3e^{j\omega_1t} \cdot e^{j\omega_1t} \cdot e^{-j(\omega_2-\Omega)(t-\Delta t)}.
\end{equation}
The second process is a non-degenerate FWM (NDFWM) process, where the generated idler pair is a copy of the non-delayed optical signal. Neglecting dispersion, the complex field of the upper NDFWM idler sideband is
\begin{equation}
E_\text{NDFWM}(t) \propto 6e^{j\omega_1t} \cdot e^{j(\omega_1 + \Omega)t} \cdot e^{-j\omega_2(t-\Delta t)}.
\end{equation}
The total sideband idler field is therefore a coherent sum of two separate idler fields with a differential delay of $\Delta t$, resulting in an interference effect. The total upper idler sideband is given by
\begin{align}
	E_\text{idler}(t) &= E_\text{DFWM}(t) + E_\text{NDFWM}(t) \nonumber\\
	&\propto (3e^{-j\Omega\Delta t} + 6)e^{j(2\omega_1-\omega_2+\Omega)t}.
\end{align}
A similar expression can be derived for the lower idler sideband.

Using an optical bandpass filter (BPF) the total idler field (idler ``carrier'' and sidebands) can finally be isolated, and its power measured with an optical power meter (i.e. low-speed photodetector) to be:
\begin{equation}
P_\text{idler} = A + B\cos(\Omega\Delta t),
\label{eq:idler_power}
\end{equation}
for some constants $A$ and $B$, with $A > B$. This expression clearly shows that once the system has been characterized, i.e. $A,B$ and $\Delta t$ are known, an optical power meter can be used to estimate the unknown RF frequency $\Omega$, without the need for expensive high-speed RF equipment.

An inspection of Eq. (\ref{eq:idler_power}) reveals the inherent trade-off between measurement bandwidth ($\propto 1/\Delta t$), and measurement sensitivity, defined as $dP_\text{idler}/d\Omega$ ($\propto \Delta t$). To bypass this trade-off and maximize measurement sensitivity without compromising bandwidth, it is necessary to maximize the idler power, both to operate further above the noise floor, and to increase the slope of Eq. (\ref{eq:idler_power}). It is then crucial to harness FWM in a low-loss, nonlinear platform.

\section{Experiment}
\subsection{Device Description}
In this work, the nonlinear platform used to harness the FWM was a 35 cm long silicon strip waveguide \cite{cherchi:13}. Silicon is a highly attractive platform for nonlinear optics partly due its large Kerr coefficient (100 times larger than that of silica), and due to its compatibility with CMOS processes \cite{leuthold:10, chen:13}.

\begin{figure}[htbp]
	\centering
	\includegraphics[width=\linewidth]{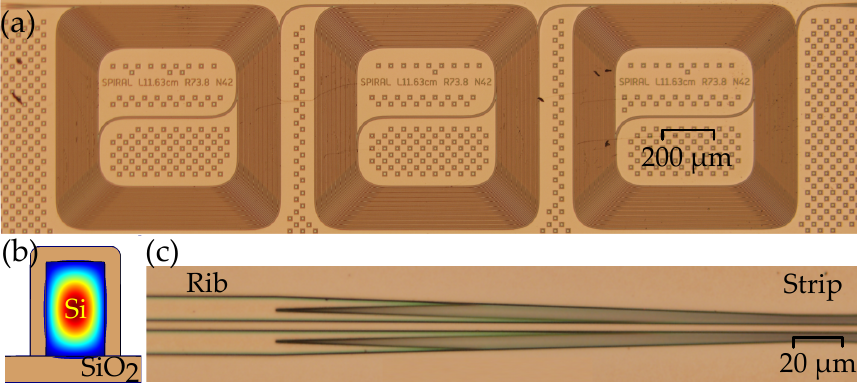}
	\caption{(a) 35 cm long, thick silicon spiral top view. (b) Simulation of the fundamental mode for the 3$\times$1.875 $\mu$m silicon strip waveguide. (c) Rib-to-strip converter for coupling to the fundamental mode.}
	\label{fig_spiral}
\end{figure}

The distinguishing feature of this waveguide was its large mode area of 2.7 $\mu$m$^2$, combined with a small footprint. Ordinarily, waveguides with such large mode areas required mm or even cm-scale bending radii, due to high mode coupling losses, and were thus constrained to lengths of a few centimeters \cite{taeed:04}. This particular sample however used Euler bends, where the bending radius continuously varies along the whole bend, ensuring minimum coupling to higher order modes \cite{cherchi:13}. This resulted in $\mu$m-scale bending radii, comparable to that of nanowires. Consequently, the whole 35 cm long spiral, shown in Fig. \ref{fig_spiral}(a), occupied less than 3 mm$^2$ surface area on an silicon-on-insulator (SOI) chip.

The dimensions of the strip cross section were 3$\times$1.875 $\mu$m. There are several advantages that follow from this large size: low 0.15 dB/cm linear propagation loss (compared to 1-3 dB/cm for silicon nanowires), a higher nonlinear loss threshold, and efficient coupling to lensed fibers (1.5 dB/facet coupling losses). The coupling was done into a single-mode rib waveguide which then tapered into the strip waveguide, as shown in Fig. \ref{fig_spiral}(c), ensuring excitation of the fundamental mode. The waveguide dispersion at 1550 nm was normal, which led to a FWM 3-dB bandwidth of 6 nm. This is sufficient for most RF photonic applications, where the signal content is in the range of 1-100 GHz, or less than 2 nm. 

The nonlinear optical properties of the waveguide were measured through a series of self-phase modulation (SPM) and FWM experiments. By launching picosecond pulses into the waveguide, we observed SPM broadening and were able to estimate the Kerr coefficient as $n_2 \sim 1.2\times 10^{-18}$ m$^2$W$^{-1}$. This value was slightly lower than for a typical silicon waveguide, possibly due to the change in crystal orientation that occurs in the spiral bends \cite{boyraz:04}. The nonlinear parameter was then estimated as $\gamma \sim$ 1.8 m$^{-1}$W$^{-1}$. We note that the maximum FWM conversion efficiency for this waveguide was comparable to that obtainable using a much shorter 5 mm standard silicon nanowire. However, our thick silicon spiral was not optimized for FWM. More importantly, the low coupling losses meant that the total waveguide insertion loss was lower than for a nanowire. This property is central to IFM applications, where the output idler power has a direct effect on the system performance.

\subsection{IFM Experiment}
The experimental setup was based on the structure shown in Fig. \ref{fig_principle}(a). Two semiconductor laser diodes were used to generate the optical carriers, with wavelengths of 1550.0 and 1551.7 nm. The electro-optic modulator (EOM) used was a Mach-Zehnder modulator, while the $\Delta t$ delay was adjusted using a tunable optical delay line.

\begin{figure}[htdp]
	\centering
	\includegraphics[scale=1]{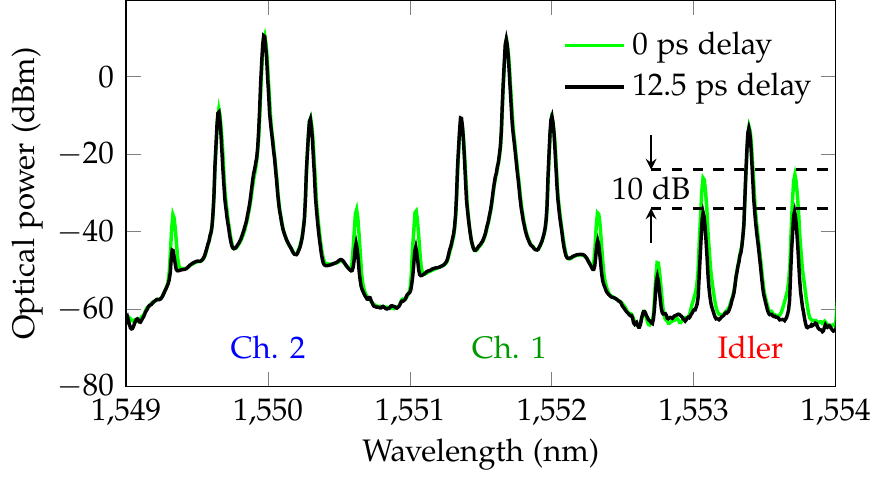}
	\caption{Optical spectra at the output of the silicon waveguide for two different $\Delta t$ values. The 10 dB power difference between the idler sidebands is a manifestation of the interference effect between the idlers generated through DFWM and NDFWM.}
	\label{fig_osa}
\end{figure}

Figure \ref{fig_osa} shows the optical spectrum at the output of the silicon spiral, when the microwave input frequency is $\Omega/2\pi=40$ GHz. By adjusting the tunable delay line, we were able to vary the product $\Omega\Delta t$ between 0 and $\pi$. According to Eq. (\ref{eq:idler_power}), these two points corresponded to maximum and minimum idler sidebands power, respectively. This is clearly visible in Fig. \ref{fig_osa}, where there is a 10 dB power contrast for the idler sidebands as the delay is tuned.
\begin{figure}[htdp]
	\centering
	\hspace{-0.15cm}
	\includegraphics[scale=1]{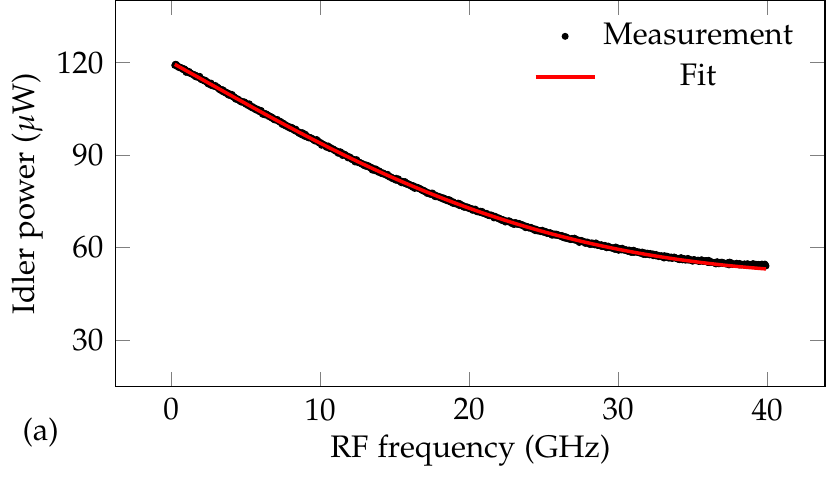}
	\includegraphics[scale=1]{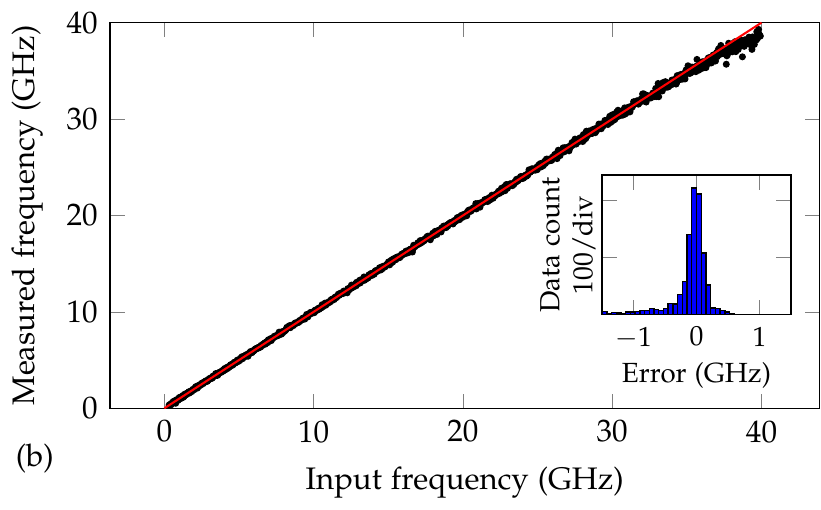}
	\caption{(a) IFM RF system response with $\Delta t =$ 8.3 ps. (b)  Frequency estimation measurement over a single 40 GHz frequency band (inset: histogram of the frequency measurement error, rms value = 318.9 MHz).}
	\label{fig_long}
\end{figure}

Initially, the time delay between the optical signals was set to 8.3 ps. Both optical carriers entered the waveguide with 18.7 dBm power, which resulted in a FWM conversion efficiency of $-$30.1 dB, here defined as $P_\text{idler,out}/P_\text{signal,in}$. The IFM system was then characterized by sweeping an RF signal generator from 0 to 40 GHz, and measuring the generated idler power. This measurement is shown in Fig. \ref{fig_long}(a), together with a theoretical fit. The fit was then used to estimate unknown frequencies in the measurement range, and the results are shown in Fig. \ref{fig_long}(b). The root mean square (rms) of the frequency estimation error was 318.9 MHz, which corresponded to 0.8\% of the 40 GHz measurement bandwidth. This combination of wide measurement bandwidth (limited only by the range of the RF signal generator), and sub-1\% error, is a record for an IMWP IFM system.
\begin{figure}[htdp]
	\centering
	\includegraphics[scale=1]{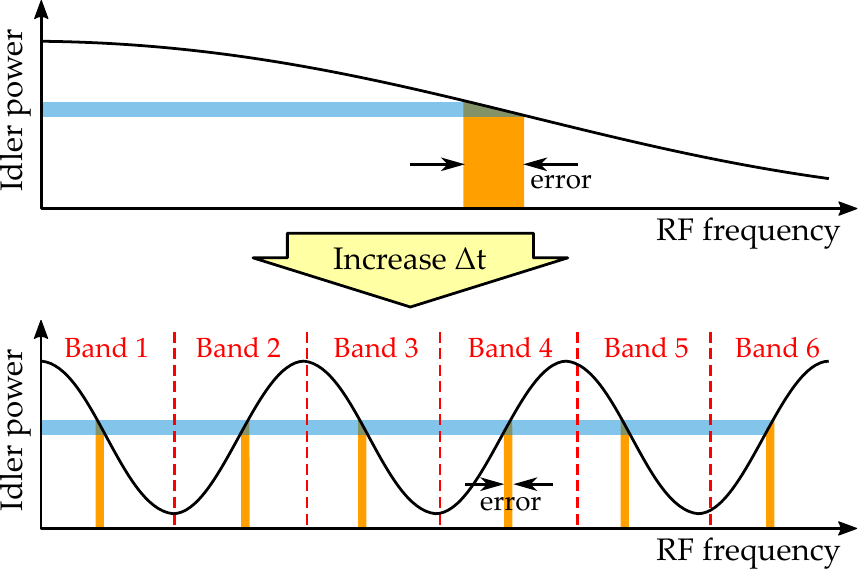}
	\caption{Reconfiguration of the IFM system response between high bandwidth/error state (low $\Delta t$) and low bandwidth/error state (high $\Delta t$).}
	\label{fig_reconfiguration}
\end{figure}

As explained in Section 2, increasing the $\Delta t$ delay can lead to enhanced measurement sensitivity (i.e. lower error), at the cost of a reduced measurement bandwidth. This trade-off is displayed in Fig. \ref{fig_reconfiguration} where, as $\Delta t$ is increased, a given idler power measurement no longer maps to a single RF frequency band, but to multiple, narrower bands. Nevertheless, by rapidly reconfiguring $\Delta t$, it is possible to exploit the high accuracy of a high-$\Delta t$ measurement, combined with the wide bandwidth of a low-$\Delta t$ measurement. This is a process consisting of two steps. Initially, a small $\Delta t$ is chosen to obtain a rough estimate of the frequency band in which the RF signal resides (e.g. Band 4 in Fig. \ref{fig_reconfiguration}). Following this, $\Delta t$ is increased, producing a highly accurate, unambiguous estimate of the signal frequency.

\begin{figure}[htdp]
	\centering
	\hspace{-0.15cm}
	\includegraphics[scale=1]{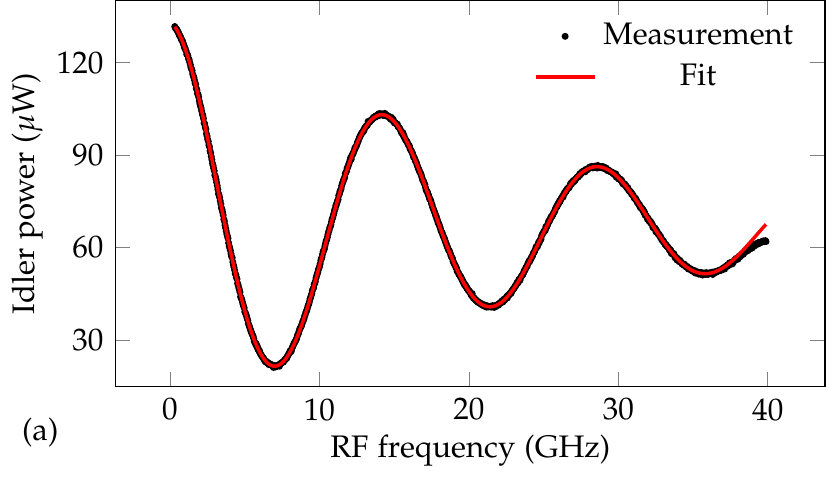}
	\includegraphics[scale=1]{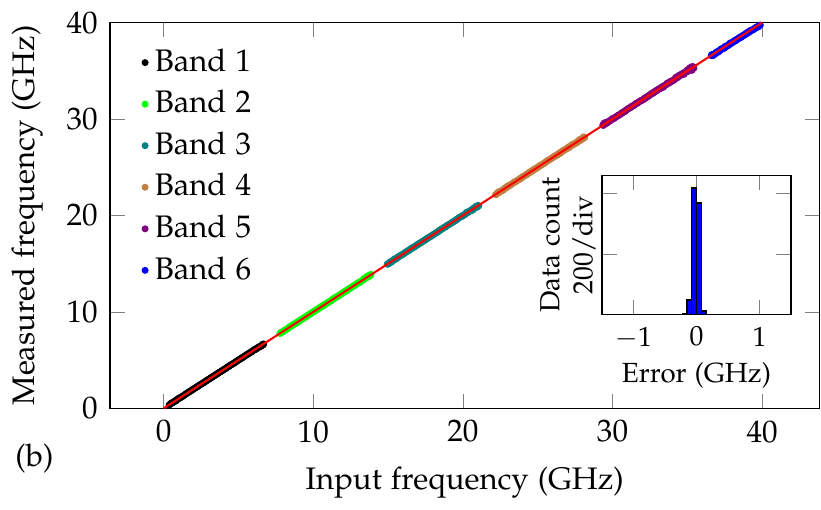}
	\caption{(a) IFM system response with $\Delta t =$ 69.4 ps. (b) Frequency estimation measurement for six separate 7.2 GHz frequency bands (inset: histogram of the frequency measurement error, rms value = 40.2 MHz).}
	\label{fig_short}
\end{figure}

To demonstrate this process, we reconfigured the system by increasing the time delay to 69.4 ps. The system characterization and theoretical fit are shown in Fig. \ref{fig_short}(a). The decaying sinusoid response is due to the frequency roll-off of the modulator. Each linear region of the system response was then assigned to a particular frequency band. This resulted in six, 7.2 GHz bands, which were then used to estimate the frequency of various RF tones. The result of this measurement, shown in Fig. \ref{fig_short}(b), exhibits a low estimation error, with rms value of 40.2 MHz, or 0.56\%  of the 7.2 GHz measurement bandwidth.

These measurements show a unique and important feature of FWM-based IFM: the ability to easily tune the system response to optimize for bandwidth or sensitivity. Therefore, it is crucial to be able to quickly tune the $\Delta t$ delay, so as to perform fast consecutive measurements with wide bandwidth and high accuracy. Our technique for implementing the delay constitutes a significant improvement over that used in previous FWM-based IFM \cite{bui:09, bui:13}, where a dispersive element introduced a delay between the two channels. This is because a tunable delay line offers a much faster mechanism for tuning $\Delta t$, compared to tuning a laser wavelength. Furthermore, a recent breakthrough in setting tunable time delay has achieved sub-nanosecond settling time \cite{bonjour:15}. This opens the way to achieving the first IFM system which overcomes the bandwidth/error trade-off through sub-nanosecond reconfiguration.

\section{Future Vision}
The ability to harness nonlinear optical processes efficiently, in a low-loss integrated platform, is fundamental for a wide variety of microwave photonic applications \cite{marpaung:13}. This is particularly true of FWM-based IFM where, as we have seen, the generated idler power has a direct effect on the measurement error. In this work, we have shown that long, thick silicon waveguides exhibit all of these properties, while being able to maintain low footprints and a high nonlinear loss threshold.

Nevertheless, to be able to monolithically integrate all critical components of the IFM system, one will need to combine the thick silicon platform with a more standard SOI technology. One such technology is 220 nm thin SOI, which enables access to a full library of active and passive components.

\begin{figure}[htdp]
	\centering
	\includegraphics[scale=1]{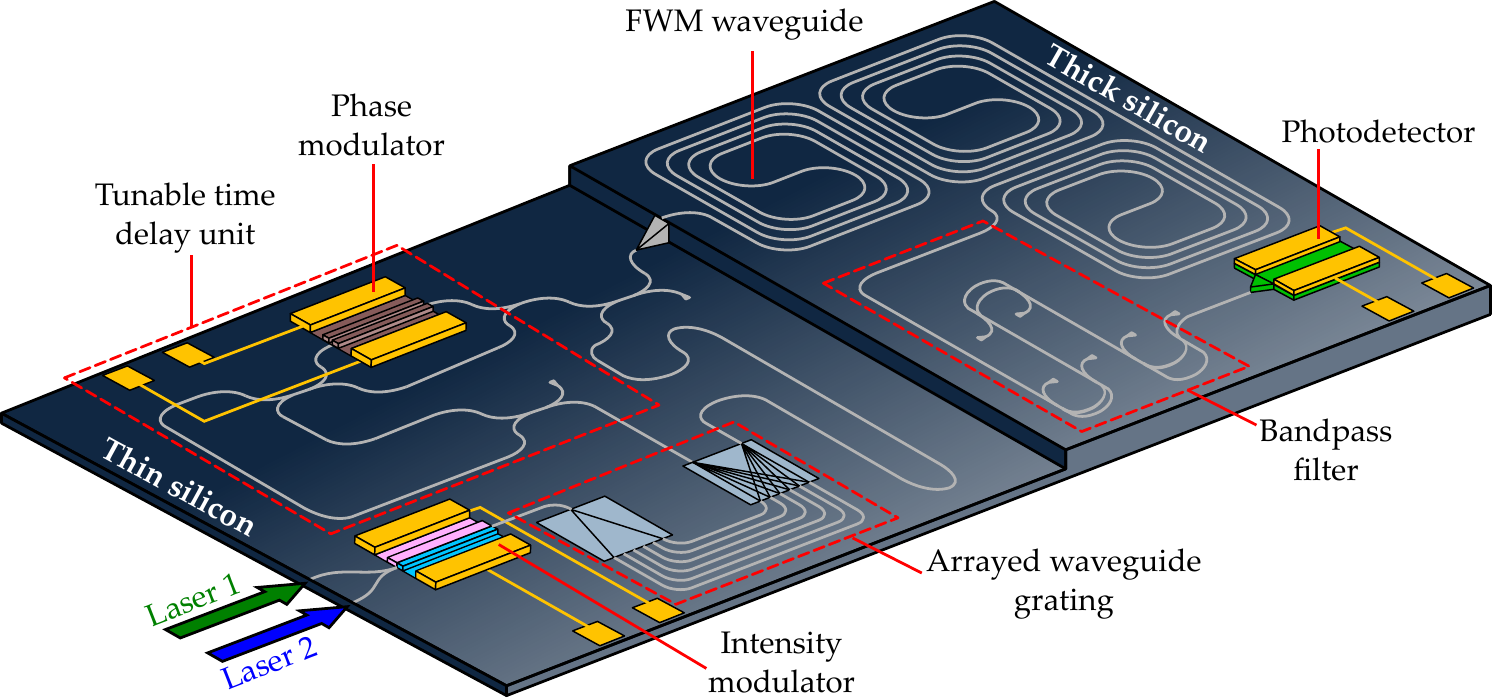}
	\caption{Future vision of the SOI integrated FWM-based IFM system. The chip is divided into thin and thick silicon sections. The RF input signal is applied to the intensity modulator. The tunable time delay unit consists of an optical delay interferometer and a phase modulator \cite{bonjour:15}. Tuning of $\Delta t$ is achieved by varying the voltage input to the phase modulator. The photodetector outputs a DC voltage proportional to the idler power, which is then mapped to an estimate for the RF input signal frequency.}
	\label{fig_future}
\end{figure}

Our vision of the layout of a future fully-integrated SOI FWM-based IFM system is presented in Fig. \ref{fig_future}, featuring a single transition from thin to thick silicon. This transition could be implemented using a section of tapered rib waveguide that minimizes the excitation of higher order modes \cite{soref:91}, or even through photonic wire bonding \cite{lindenmann:12}.

The thick silicon part of the chip houses the FWM waveguide, BPF, and photodetector. One of the main advantages of the current FWM-based IFM scheme is that it requires only a simple low-speed photodetector. Such a device has already been demonstrated using vertical p-i-n germanium, integrated on a thick SOI waveguide \cite{feng:10}. Bandpass filters composed of cascaded Mach-Zehnder interferometers have also been demonstrated in thick silicon \cite{cherchi:14}. For these components, Euler bends will need to be implemented so as to maintain a low footprint.

The thin silicon section of the chip hosts the electro-optic modulators, and arrayed waveguide grating. The intensity modulator used by the input RF signal would be implemented using doped thin SOI technology. Such Mach-Zehnder modulators have been shown capable of high performance, with low 1.6 dB insertion loss, and wide 27.8 GHz bandwidth \cite{xiao:13}. Demultiplexing of the two channels would occur in an arrayed waveguide grating, which have also been demonstrated in thin SOI with 200 GHz channel spacing \cite{pathak:14}. Finally, tunable time delay could be realized using a novel approach \cite{bonjour:15}, employing an optical delay interferometer and a phase modulator. High-speed phase modulators have been implemented using silicon-organic hybrid (SOH) technology \cite{alloatti:14}, and would allow for fast tuning of the $\Delta t$ delay. Such a chip, boasting fast tunable delay and compact size will result in the first ultra-low latency and highly accurate IFM system with bandwidth and capabilities beyond what is achievable using state-of-the-art RF technologies.
 
\section{Conclusion}
We have presented the first IFM system using on-chip FWM, capable of extremely high frequency measurements, and low error through easy reconfiguration of the system response. The enabling technology was a low-loss, long silicon waveguide, which was used to generate strong FWM idlers. This allowed us to achieve record-low measurement error over a wide frequency range, greatly surpassing that of previous IMWP IFM systems. The novel setup we have presented consists of components which are all capable of SOI integration, opening the way for the first reconfigurable, monolithically integrated, IFM receiver.

\section*{Funding Information}
Australian Research Council (ARC) (DE150101535, FL120100029, CE110001018).

\end{document}